May 17, 2006

# On connection between classical and quantum mechanics

Daniel Sepunaru

RCQCE - Research Center for Quantum Communication,

Holon Academic Institute of Technology,

52 Golomb St., Holon 58102, Israel

**Abstract**

We discuss an alternative version of non- relativistic Newtonian mechanics in terms of a real Hilbert space mathematical framework. It is demonstrated that the physics of this scheme correspondent with the standard formulation. Heisenberg-Schrödinger non-relativistic quantum mechanics is considered adequate and complete. Since the suggested theory is dispersion free, linear superposition principle is not violated but cannot affect results of measurements due to spectral decomposition theorem for self-adjoint operators (the collapse of wave function).

**Introduction.**

The purpose of this investigation is to establish for classical mechanics the structural framework similar to the one used in quantum theory. We restrict ourselves to description of single particle states and prefer here to avoid complications introduced by special relativity. In order to make clear the mathematical and correspondent physical content of successive discussion, I will quote the following statement [1]:

If $\hat{A}^+ = \hat{A}$ and $<\Psi|\Psi_1> = 0$, we can always decompose

$$\hat{A}|\Psi> = \alpha|\Psi> + \beta|\Psi_1>$$

with $\beta$ real and non-negative.

$$\alpha = <\hat{A}> \equiv \overline{A}$$

$$\beta = \left[<\hat{A}^2> - (<\hat{A}>)^2\right]^{1/2} \equiv \Delta A.$$

This theorem appears several times [2] in different contexts, but in form presented, its content takes a clear view of the situation: it is enough that wave function of the system will contain two linearly independent (orthogonal) components in order that the correspondent observable will have non-zero dispersion. The non-zero dispersion leads to practically instant spread of wave packet.

Now, let us consider famous E. Schrödinger cat example [3]. The essential points are: 1) The cat may be presented as a quantum mechanical system and not as a classical measurement instrument; 2) The system state is described by the following linear superposition of pure states:

$$|\Psi_{cat}> = \frac{1}{\sqrt{2}}(|\Psi_{alive}> + |\Psi_{dead}>).$$

E. Schrödinger did not continue discussion after that point. But since a cat is in the superposition state, this will lead to the spread of wave packet within time uncertainty predicted by W. Heisenberg. The curious experimenter will find cat "blurred" over entire volume of the chamber and disappeared (from classical point of view) together with his smile (notice that if it was correct, then the quantum mechanics would provide proper unification between L. Carroll and E. Schrödinger fantasies). It is remarkable that E. Schrödinger concluded discussion of cat paradox by the following statement:

"It is typical of these cases that an indeterminacy originally restricted to the atomic domain becomes transformed into macroscopic indeterminacy, which can then be *resolved* by direct observation".

That statement is in contradiction with the J. von Neumann conjecture [2] that the macroscopic physics are dispersion free.

**Real Hilbert Space Formulation of Classical Mechanics**

First of all I should define what I mean by real Hilbert space. We should maintain the connection with quantum theory as close as possible and assure the proper extension to relativistic version.

In addition, a scheme should incorporate classical electrodynamics through application of principle of local gauge invariance. Therefore we will use the following definitions:

| Real Hilbert space | Complex Hilbert space |
|---|---|
| System state: complex wave function | System state: complex wave function |
| Dynamical variable: complex linear operator | Dynamical variable: complex linear operator |
| Observable: self-adjoint (hermitian) operator | Observable: self-adjoint (hermitian) operator |
| Measurement of observable value [*]: $<\psi \mid \hat{A}_{clas} \mid \psi> = tr\int \bar{\psi} \hat{A} \psi d^3x$ | Value of observable: $<\psi \mid \hat{A}_{quant} \mid \psi> = \int \bar{\psi} \hat{A} \psi d^3x$ |

[*] Scalar product in this framework is defined by:

$$<f \mid g>_R = tr\int \bar{f} g d^3x,$$

with underlined numerical basis of dimension two (complex numbers). This implies that

$$x^2 - tr(x)x + N(x)1 = 0 \quad \forall x = complex, \; tr(x) = real, \; N(x) = real$$

$$x + \bar{x} \equiv tr(x)1, \quad x\bar{x}(= \bar{x}x) \equiv N(x)1.$$

In particular,

$$tr(i) = tr\begin{pmatrix} 0 & -1 \\ 1 & 0 \end{pmatrix} = 0.$$

Notice that in quantum theory the relevant scalar products associated with observable quantities are always real. Since in classical mechanics every dynamical variable is observable, we will further discuss only self-adjoint operators. They satisfy the following algebra:

1) $\hat{C} \equiv \hat{A} + \hat{B}, \quad \hat{C} = \hat{C}^+ \quad \forall \hat{A}, \hat{B} \quad \hat{A} = \hat{A}^+, \hat{B} = \hat{B}^+.$

This statement will be justified after introduction of a product of two self-adjoint operators.

2) $\hat{C} \equiv \hat{A}\hat{B}, \quad \hat{C} = \hat{C}^+ \quad \forall \hat{A}, \hat{B} \quad \hat{A} = \hat{A}^+, \hat{B} = \hat{B}^+.$

Proof.

$$\hat{A}\hat{B} = \frac{\hat{A}\hat{B} + \hat{B}\hat{A}}{2} + \frac{\hat{A}\hat{B} - \hat{B}\hat{A}}{2} = \frac{\hat{A}\hat{B} + \hat{B}\hat{A}}{2} + i\hat{D}$$

$$\hat{B}\hat{A} = \frac{\hat{A}\hat{B} + \hat{B}\hat{A}}{2} + \frac{\hat{B}\hat{A} - \hat{A}\hat{B}}{2} = \frac{\hat{A}\hat{B} + \hat{B}\hat{A}}{2} - i\hat{D}$$

Obviously, $\hat{D} = \hat{D}^+$.

Then

$$< \hat{A}\hat{B} >_R = < \frac{\hat{A}\hat{B} + \hat{B}\hat{A}}{2} >_R = tr < \hat{A}\hat{B} >_C$$

$$< \hat{B}\hat{A} >_R = < \frac{\hat{A}\hat{B} + \hat{B}\hat{A}}{2} >_R = tr < \hat{B}\hat{A} >_C$$

Therefore

$$< (\hat{A}\hat{B} - \hat{B}\hat{A}) >_R = 0$$
$$\hat{A}\hat{B} = \hat{B}\hat{A}$$

with respect to real scalar product defined above.

Since the product of any pair of self-adjoint operators is a self-adjoint operator, the product of an arbitrary number of self-adjoint operators is a self-adjoint operator.

Now let us examine properties of self-adjoint operators in these schemes.

1. Spectral Decomposition Theorem: For every self-adjoint (hermitian) operator $\hat{A}$ that is suitable for description of the observable dynamical variable (has all necessary properties required by functional analysis), there exists the complete orthonormal basis in Hilbert space uniquely defined by the requirement

$$\beta = \left[ <\hat{A}^2> - (<\hat{A}>)^2 \right]^{1/2} \equiv \Delta A = 0 \tag{1}$$

The components of that basis are the solutions of the equation

$$\hat{A} f_k = \lambda_k f_k \tag{2}$$

which is a consequence of eq. (1). $f_k$ are called eigenfunctions of the operator $\hat{A}$. Sets of real numbers $\lambda_k$ are called eigenvalues of the operator $\hat{A}$, are defined simultaneously with $f_k$ and form a spectrum of $\hat{A}$. The spectrum of hermitian operator may contain several discrete numbers, a countable set of discrete numbers or/and continuous interval (finite or infinite).

The careful reader may verify line by line that there is no difference between complex and real Hilbert spaces as defined above with respect to spectral decomposition theorem. There exists vast literature on the topic but the books of R.Courant and D.Hilbert [4] and J.von Neumann [2] still remain useful.

Here perhaps I should add the important remark. When we write

$$\hat{A} f_k = \lambda_k f_k,$$

we usually say that it provides physical value(s) of the observable $\hat{A}$. It does not. Wave function $f$ is not an observable quantity. Wave function $g = \hat{A} f$ is also an unobservable quantity. The value of an observable should be a real number. Only the expression

$$<\psi | \hat{A}_{quant} | \psi> = \int \overline{\psi} \hat{A} \psi d^3 x$$

provides the values of the observable in quantum theory. Therefore, the additional relation is required in order to associate them with the results of measurements.

2. The necessary and sufficient condition for two or any number of hermitian operators to have a common set of eigenfuctions which form a complete orthonormal basis in Hilbert space is that they are mutually commuted. Since the product of two or any number of mutually commuting hermitian operators is again a hermitian operator (and commutes with each of it components), it has the same set of eigenfunctions. Indeed, every one of them in that basis is dispersion free.

Hence, the real Hilbert space as defined above provides realization of dispersion free physical theory.

Moreover, since the coordinate $\hat{x}$ has a purely continuous spectrum, every observable in that theory has a continuous spectrum. An additional feature, which distinguishes it from the complex Hilbert space framework, is uniqueness of its basis. The theory remains linear and does not exclude validity of linear superpositions for the system states; however, only precise values of hermitian operators are measured. That phenomenon is known as collapse of wave function.

3. Another mathematical statement that may have a very interesting physical realization (we will discuss it later) is valid in real Hilbert space: for an arbitrary set of mutually commuting hermitian operators $\hat{A}, \hat{B}, \hat{C},...$ there exists hermitian operator $\hat{R}$ such that each one of $\hat{A}, \hat{B}, \hat{C},...$ is a function of $\hat{R}$ [5].

Now it become manifestly obvious that real Hilbert space provides a convenient arena for Newtonian mechanics. Finally, let us demonstrate that the equation

$$<\varphi|\hat{B}_{clas}|\varphi> = <\psi|\hat{B}_{quant}|\psi> \qquad (3)$$

is equivalent to the Heisenberg quantization condition.

It is well known that the classical equations of motion

$$\dot{q}_i = \frac{\partial H}{\partial p_i}, \quad \dot{p}_i = -\frac{\partial H}{\partial q_i}$$

have the following form in terms of hermitian operators [6]:

$$\frac{d\hat{A}}{d\hat{t}} = \frac{\partial \hat{A}}{\partial \hat{t}} + \sum_i (\frac{\partial \hat{A}}{\partial \hat{q}_i} \frac{\partial \hat{H}}{\partial \hat{p}_i} - \frac{\partial \hat{A}}{\partial \hat{p}_i} \frac{\partial \hat{H}}{\partial \hat{q}_i})$$

if we choose

$$\hat{B}_{clas} = \sum_i (\frac{\partial \hat{A}}{\partial \hat{q}_i} \frac{\partial \hat{H}}{\partial \hat{p}_i} - \frac{\partial \hat{A}}{\partial \hat{p}_i} \frac{\partial \hat{H}}{\partial \hat{q}_i})$$

and

$$\hat{B}_{quant} = \frac{i}{\hbar}(\hat{H}\hat{A} - \hat{A}\hat{H})$$

then using eq. (3) we obtain

$$\frac{d\hat{A}}{dt} = \frac{\partial \hat{A}}{\partial t} + \frac{i}{\hbar}(\hat{H}\hat{A} - \hat{A}\hat{H})$$

which are the quantum equations of motion written in the Heisenberg representation.

The eq. (3) is the fundamental relation that defines the results of measurements. For the model example of a particle in an infinite spherical well, only discrete solutions of right hand side imbedded into continuous spectrum of left hand side will be revealed.

**Theory of measurements**

In the previous section the formulation of Newtonian dynamics was achieved and was even demonstrated that the same dynamical law still governs time evolution of the system in quantum physics. It was discovered by E. Schrödinger [7] that the alternative equivalent form of equations of motion exists in non-relativistic version of the theory. However, what that which is mentioned by the notion "physical law" still remains undefined.

We may say that for the mathematical structure it is sufficient to be legal and legitimate if its foundation is based on mutually consistent set of assumptions. It is not sufficient for physics: physics are an empirical science (in practice this means that the only perfect mathematical frameworks will survive). The realization of that requirement is performed through introduction of properly defined measurement instruments and procedures. The structure of Newtonian mechanics represents the basic ingredients needed to achieve that. In addition to the formulation of the time evolution of the system (dynamics), the equations of motion should earn status of physical law. The latter requirement is satisfied by the introduction of the reference frames and rules as how they are connected with each other. It is meaningless to discuss any law or relation without its universality with respect to a chosen and well defined infinite set of reference frames. In the classical mechanics, it turns out that the definition of the suitable reference frame (inertial systems) occurs through the idealization of the free moving body isolated from it environment. Then the connection between those frames is given in terms of the motion of such a body. This is the content of the first law of the Newtonian mechanics (Galileo law of inertia), which is indeed consistent with the fundamental equations of motion. We associate that body, located at the origin of a given reference frame, with an appropriate set of measurement instruments. Therefore the origin of the reference frame should be defined with certainty and the measurement instruments should obey laws of classical physics.

In the scheme suggested here, both requirements are met and eq. (3) assures that all relevant information about properties of investigated quantum mechanical system is available. Eq. (3) plays a role analogous to the third Newton law.

A single isolated sample of the experimental data has no meaning in classical physics. Only repetitions of the sample will confirm that the obtained result represents the objective reality. The requirement that the system state remains unchanged during the experiment was never fulfilled; even the system invariants (for example in collisions) may change. What is essential is that if the consecutive (in time) measurements on the same system are not legal, the repetition of the measurement should be assured by possibility to prepare a system identical to the original one. As pointed out by E. Schrödinger [3] the measurements of quantum mechanical systems allow unprecedented precision as well as preparation of the identical experiments (feature inherent to objective property of the quantum systems- the identical particles are undistinguishable; a situation which is not available within the classical world).

The collection of the obtained results is now the subject of the standard techniques for the data processing. In case the system under test is a classically defined material point (coordinate and momentum are mutually commuted hermitian operators), one will obtain a picture sharply concentrated around a single isolated point. In case when the system obeys the laws of quantum physics, one will get a picture of a spatially extended object; the number of required samples is determined by the classical methods of image and/or signal processing [8]. I do not see any importance to our subject that the sequence of samplings emerged in random fashion. There is no doubt that the obtained result reconstructs the objective reality exactly in the same manner as the image of the classical material point was obtained.

In the present discussion, we restrict consideration to the non-relativistic limit of classical and quantum physics. Inclusion of electromagnetic interaction (as well as gravitation) leads to a relativistic local field theory. No measurement instruments or procedures exist that violate foundations of classical physics. Contradictions currently discussed in literature are apparently only a matter of misinterpretation.

**Conclusions**

It was demonstrated that non-relativistic classical mechanics might be reformulated in terms of real Hilbert space, with an underlined numerical system of complex numbers. It is worthwhile to mention that similar structures based on real quaternions and real octonions exist [9]. The presence of rich phase structure in the definition of wave functions (system states) should allow the axiomatic introduction of electromagnetic and gravitational interactions by means of application of the principle of local gauge invariance. The relativistic version of the theory is expected to emerge naturally in suggested frameworks.

The present paper is devoted mainly to the problems related to the measurement theory. The beautiful books written by R. Penrose [10] inspired my investigation. However, the results presented apparently do not support the ideas developed in it. Perhaps, it is not so. Within a classical world we are working in Heisenberg representation. Hilbert space appears to be uniquely defined and rigid and plays a role of passive arena for the events associated with the dynamics of the physical system. The space-time continuum plays a similar role in the standard formulation of Newtonian mechanics. It seems to me reasonable to expect that these arenas are actually identical. H.Weyl [11] developed a technique suitable for verification of this conjecture and the J.von Neumann theorem mentioned above (statement 3) may assign to it the dynamical content.

We have used here one-particle states only and discussed the role of dispersion defined as the property of the operators acting in Hilbert space formed by that state. The origin of the terminology lies in the statistical interpretation of quantum physics. Unfortunately, I was never able to understand the arguments behind this interpretation. So, may be it is right, may be it is not adequate. I used it in order not to confuse the reader similarly to as we say: "Tomorrow sun will rise at six thirty". A more disturbing feature of statistical interpretation is that it probably rejects the possibility to gain precise knowledge of the physical systems in quantum world. J.M.Jauch [12] describes that process as a journey in the infinite library, which contains all the answers. The journey only starts. Above the entrance door it is written: "Igitur eme, lege, fruere".

I am grateful to I.D.Vagner and L. Sepunaru for discussions.